\documentstyle[prb,aps,twocolumn]{revtex}
\input epsfig.sty      
\begin{document}

\advance\textheight by 0.2in 
\twocolumn[\hsize\textwidth\columnwidth\hsize\csname@twocolumnfalse\endcsname
\draft

\begin{flushright}
{\tt to appear in Phys. Rev. B}
\end{flushright}

\title{Phase-coherence threshold and vortex-glass state in diluted 
Josephson-junction arrays in a magnetic  field}
\author{Enzo Granato}
\address{Laborat\'orio Associado de Sensores e Materiais,
Instituto Nacional de Pesquisas Espaciais, 12201-190 S\~ao Jos\'e 
dos Campos, SP Brazil}
\author{Daniel Dom\'{\i}nguez}
\address{ Centro At\'{o}mico Bariloche, 8400 San Carlos 
de Bariloche, Rio Negro, Argentina}

\maketitle

\begin{abstract}
We study numerically the interplay of
phase coherence and vortex-glass state in  two-dimensional Josephson-junction 
arrays with average rational values of flux quantum 
per plaquette $f$ and random  dilution of junctions. 
For $f=1/2$, we find evidence of a phase coherence threshold value $x_s$,
below the percolation concentration of diluted junctions $x_p$, 
where the superconducting transition vanishes. For $x_s < x < x_p$ the 
array behaves as a  zero-temperature vortex glass 
with nonzero linear resistance at finite temperatures. The  zero-temperature   
critical currents are insensitive to variations in $f$ in the vortex glass
region while they are strongly $f$ dependent in the phase coherent region.  
\end{abstract}


]

The study of the structure of vortex-lattice states in disordered
superconductors in a magnetic field have attracted much recent interest. In
three-dimensions a true vortex-glass transition \cite{fisher} at finite
temperature is possible for strong disorder in unscreened superconductors 
\cite{by}, while for weak disorder  a Bragg-glass phase with quasi-long range
order has been proposed \cite{gld}. In two-dimensions, however, vortex-glass
models \cite{hyman,eg98} and experiments on superconducting films \cite{dekker} 
show that vortex-glass order and phase coherence are destroyed at
any finite temperature with a nonzero but exponentially small resistivity in the
large disorder limit whereas in the  weak disorder limit the situation is
less clear and may depend on the particular model of disordered superconductor.
Random diluted Josephson-junction arrays have been used to model disordered
superconductors \cite{john,stroud,li91,gd97,bgsg} 
and can also be fabricated with controlled
amount of disorder in two-dimensions \cite{harris,martinoli,affolter}. For a
regular array in perpendicular magnetic field with a rational flux quanta
per cell $f$, the ground sate consists of a periodic pinned vortex lattice,
with additional discrete symmetries resulting from commensurability effects 
\cite{conf,teitel}, and phase coherence and vortex order is possible. 
Thus, diluted arrays in a magnetic field can provide a convenient experimental 
model system to investigate the effects of {\it weak} and {\it strong} disorder 
on initially pinned vortex lattices and the interplay of phase coherence and vortex glass
states in two dimensions. In particular, in order to understand transport
properties near percolation threshold in recent experiments \cite{affolter}
on diluted arrays in a magnetic field, it is important to know if disorder
and temperature fluctuations can destroy phase coherence at long length
scales and the nature of vortex order in this regime. 
A recent study \cite{gt99} of a model of random Josephson-junction 
arrays with a particular type of disorder (positional disorder)\cite{gk89}
in a magnetic field suggests that no transition is possible 
even for weak disorder in the thermodynamic limit but it is not clear if this
scenario would apply in general. In fact, random dilution does not explicitly
introduces random phase shifts across the junction unlike positional disorder.
In addition, an earlier study of the ground-state stability of a diluted array 
shows  \cite{bgsg} that, in presence of a magnetic field with an average rational 
value of flux quantum per plaquette $f$, phase coherence is possible at nonzero 
temperatures and the transition temperature only vanishes, 
for increasing dilution of junctions $x$, at a critical
value $x_{s}(f)$  below the geometric percolation threshold $x_{p}$
where the transition would vanish in the absence of the magnetic field. A
vortex-glass phase with zero-temperature transition should also appear at a
critical concentration $x_{v}\gtrsim x_{s}$, with nonzero resistivity at
finite temperatures but having a diverging short-range correlation length $%
\xi \propto T^{-\nu }$ which is expected to determine the nonlinear
behavior of the current voltage characteristics. An upper bound for the
phase-coherence region is set by the behavior at $f=1/2$ since higher order
rational values are expected to be much less stable with a corresponding 
threshold value which may be too small to detect numerically. 
These results rely on the finite-size behavior of 
defect energy in the ground state  which are inaccessible experimentally. 
However, experiments often measure transport
properties and it is of great interest to know how these effects could show
up in the behavior of the current-voltage characteristics.

In this work, we present the results of extensive dynamical simulations of
the current-voltage characteristics of resistively shunted
Josephson-junction arrays with an average flux quantum per plaquette $f$ and
random dilution of junctions. We find evidence of a phase coherence
threshold value $x_{s}<x_{p} $ as indicated in Fig. 1. For $x<x_{s}$, the
superconducting transition occurs at finite temperatures while for $%
x_{s}<x<x_{p}$ the array behaves as a zero-temperature vortex glass with
nonzero and thermally activated linear resistance at finite temperatures and
diverging short-range correlation length $\xi \propto T^{-\nu }$. A
current-voltage scaling analysis provides an estimate of $\nu \sim 2$. In
the vortex-glass region, the zero-temperature critical currents are roughly
insensitive to changes in $f$.

We consider a two-dimensional array of superconducting grains coupled to its
nearest neighbors by resistively shunted Josephson junctions and with
current conservation at each site \cite{shenoy}. The equations of motion for
the phases $\theta _{i}$ of the superconducting order parameter located at
site $i$ of the lattice can be written as \cite{gd97,dd99} 
\begin{equation}
\frac{\hbar}{2eR_{o}}\sum_{j}(\dot{\theta}_{i}-\dot{\theta}%
_{j})=-\sum_{j}[I_{ij}\sin (\theta _{i}-\theta _{j}-A_{ij})+\eta _{ij}]
\end{equation}
where $R_{o}$ is a uniform shunt resistance, $\eta _{ij}(t)$ is a thermal
noise with correlations $\langle\eta_{ij}(t)\eta_{kl}(t^{\prime})\rangle
=2k_{B}T/R_{o} \delta_{ij,kl}\delta(t-t^{\prime})$ and $I_{ij}$ is the
junction critical current. The bond variables $A_{ij}$ correspond to the
line integral of the vector potential and are constrained to $%
\sum_{ij}A_{ij}=2\pi f$, about each elementary plaquette of the reference
(undiluted) lattice. For simplicity we consider a square lattice array and
bond dilution of junctions. The qualitative behavior and critical exponents
presented below should remain the same for other choices of dilution and for
triangular arrays. Dilution of junctions is introduced by taking $I_{ij}=0$
with probability $x$ and $I_{ij}=I_{o}$, a constant, with probability $1-x$.
Dimensionless quantities are used with time in units of $\tau =\hbar
/2eR_{o}J_{o}$, current in units of $I_{o}$, voltages in units of $%
R_{o}I_{o} $ and temperature in units of $\hbar I_{o}/2ek_{B}$. A total
current $I$ is imposed uniformly in the array using periodic boundary
conditions \cite{dd99} with current density $J=I/L$, where $L$ is the system
size and the average electric field $E$ is obtained from the voltage $V$
across the system as $E=V/L=(\hbar/2e)\langle\dot{\theta}_{i}-\dot{\theta}%
_{j}\rangle$. We use periodic boundary conditions in order to eliminate
possible edge contributions to the resistance due to diluted junctions near
the boundary which could arise from open boundary conditions \cite{gks}. System
sizes ranging from $L=8 $ to $L=128$ were used in the calculations with a 
time step  $\Delta t=0.07\tau $ and the results averaged over $10$ to $500$  
random diluted configurations of junctions depending on the system size.

Fig. 1(b) shows the behavior of the critical current density $J_{c}$ where a
nonzero voltage appears at zero temperature. At low values of $x$, the
behavior of $J_{c\text{ }}$ strongly depends on the rational frustration $%
f=p/q$, as indicated for $f=1/2$ and $f=1/4$ in the Figure, but becomes
roughly insensitive to $f$ for dilutions larger than a critical value much
below the percolation threshold $x_{p}$. This is consistent with the
proposed vortex glass phase \cite{bgsg} for the range $x_{v}(f)<x<x_{p}$
where $x_{v}$ is a dilution threshold below which vortex-lattice order
remains. In the undiluted case and for small enough $x$ the ground state energy 
and critical currents correlate with the ordering of $q\times q$ unit cells and 
so are very sensitive to the $q$ value \cite{teitel}. However, for $x>x_v$, 
vortex-lattice order is completely destroyed at
long-length scales and its stability and therefore the critical current
should be less sensitive to $q$. Since it is expected that \cite{bgsg} $%
x_{v}(1/4)<x_{v}(1/2)$ and $x_{v}\gtrsim x_{s}$, the change in the behavior
of $J_{c}$ in Fig. 1(b) allows for a very rough estimate of the phase coherence
threshold for $f=1/2$ as the value of dilution where the two curves overlap
within the estimated errorbars, $x_{s}(1/2)\sim 0.20(5)$ .

We turn now to thermal fluctuation effects. Fig. 2(a) shows the temperature
dependence of the nonlinear resistivity $E/J$ at a value of dilution $x=0.1$
below the phase-coherence threshold $x_{s}$ estimated above, for the largest 
systems sizes $L=64$ and $L=128$. As can be seen
from the Figure, the linear resistivity $R_{L}=\lim_{J\rightarrow 0}E/J$,
estimated from the ratio $E/J$ when $J\rightarrow 0$, tends to a finite
value at high temperatures but extrapolates to very low values at lower
temperatures, independent of system size, consistent with the existence 
of a finite temperature superconducting transition in the range  
$T_{c} = 0.3$ to $0.4$. This is confirmed by a
scaling analysis of the nonlinear resistivity according to which \cite{hyman}
measurable quantities scale with the diverging correlation length $\xi $
near the transition temperature. If the transition occurs at a finite
temperature, the relaxation time diverges as $\xi ^{z}$ , where $z$ is the
dynamical critical exponent, and the nonlinear resistivity satisfy the
scaling form

\begin{equation}
T\frac{E}{J}=\xi ^{-z}g_{\pm }(\frac{J}{T}\xi )  \label{scaling}
\end{equation}
in two dimensions, where the $+$ and $-$ correspond to the behavior above
and below the transition, respectively. For a transition in the
Kosterlitz-Thouless (KT) universality class, the correlation length should
diverge exponentially as $\xi \propto \exp (b /|T/T_{c}-1|^{1/2})$, 
while for a conventional transition
a power-law behavior is expected $\xi \propto |T/T_{c}-1|^{-\nu }$, with
an exponent $\nu $ depending on the discrete symmetry of the pinned vortex
lattice. A scaling plot according to Eq. (\ref{scaling}) can be used to
verify the scaling arguments and the assumption of finite temperature 
equilibrium transition. This is shown in Fig. 2(b), in the 
temperature range closest to the apparent $T_c$ and smallest current 
densities, assuming the correlation length $\xi $ has an exponential 
divergence as in the KT universality class and using  $b$, $T_c$ and $z$ as adjustable
parameters so that the best data collapse is obtained. As shown in the
Fig. 2(b), the two largest system sizes $L=64$ and $L=128$ give the same data collapse 
and so finite size effects, ignored in the scaling form of Eq. (\ref{scaling}), 
are not dominant for this range of temperatures and current densities. 
We estimate a transition temperature $T_c=0.33(2)$ and dynamical
exponent $z=1.2(2)$. Although this estimate is based on a scaling analysis of the
nonlinear current-voltage characteristics, which is a nonequilibrium property, we
find that the finite size behavior of the linear resistance at $T_c$ is consistent with
this analysis. In a finite system the divergent correlation length $\xi$
is cut off by the system size $L$ at the transition. From Eq. (\ref{scaling}), 
the linear resistance at $T_c$ should then scale as $R_L \propto L^{-z}$. 
The linear resistance can be obtained from the Kubo formula of
equilibrium voltage fluctuations  as $R_L=(1/2T) \int d t <V(t) V(0)> $, 
without finite current effects, and can also be  determined from the long-time
fluctuations of the phase difference across the system \cite{eg98}. Fig. 3, shows the
finite size behavior of $R_L$ at different temperatures. Above our estimate of $T_c$ it
remains finite for increasing $L$ whereas below $T_c$ it appears to extrapolate to zero.
Right at $T_c$, a power-law fit gives $z=1.07(5)$ which is consistent with the estimate
from the current-voltage scaling and suggests therefore that the transition corresponds to the
underlying equilibrium behavior. 
It should be noted that for the pure KT transition a dynamical
exponent  $z=2$ is expected, independent of the particular dynamics. 
Indeed, for $f=0$ and $x=0$ the same power-law fit gives $z=2.0(1)$ at the critical temperature. 
However, for $f=1/2$, where an additional Ising order parameter is present,
it is found that even for the undiluted system $z < 2$ using the  
present dynamics \cite{dg00}. 
We also note that attempting a scaling plot using the conventional
power-law correlation length gives a very large value for $\nu$ which
suggests that the exponential form is the appropriate one. However at $x=0$, 
both forms of correlation length gives reasonable data collapse as
expected for $f=1/2$ from the single transition scenario where the
superconducting transition and vortex-lattice disordering transition occurs
at the same temperature or else at very close 
temperatures \cite{teitel,gkn,rj,olsson}. 
Since the undiluted array at $f=1/2$ is expected to
have a transition combining the KT and Ising universalities, our results
suggest that for $0\ll x <x_{s}$, the superconducting transition is in
the KT universality class (statics) while the vortex-lattice disordering transition of
Ising symmetry may occur separately in presence of weak disorder \cite{bgsg}.
However, our above scaling analysis based on the diverging phase-coherence
correlation length does not allow us a determination of the vortex-lattice
disordering transition since it is expected to occur 
within the normal phase \cite{bgsg}.
Using the above scaling analysis, the transition temperatures for 
the different dilutions can be obtained as in Fig. 1(a). 
For values of $x \ne 0.1$ 
limited data was used and the results are only  rough estimates of the
transition temperatures.  
Nevertheless, the critical temperature as a function of 
dilution reasonably  extrapolates to the threshold $x_{s}$ 
estimated from the behavior
of $J_{c}$ in Fig. 1(b) as discussed above. 

In contrast, for a dilution above the phase-coherence threshold $x>x_{s}$,
the linear resistance $R_{L}$ is finite for all temperatures in the same
range as indicated in Fig. 4a. Although we can not exclude a transition at
much lower temperatures based on these data, 
the behavior is consistent with a superconducting
transition and vortex order occurring only at zero temperature as for a
vortex glass with a zero-temperature transition \cite{hyman,eg98}. This is
consistent with defect energy calculations which show that low-energy
excitations above the ground state decreases with system size in this range
of dilutions \cite{bgsg}. In fact, $R_{L}$ decreases rapidly with decreasing
temperature and for increasing $J$ there is a smooth crossover to nonlinear
behavior at a critical current $J_{nl}$ which also decreases with decreasing
temperature. From the well-known scaling arguments \cite{hyman} leading to
Eq. (\ref{scaling}), if the transition happens only at zero temperature then 
$\xi \propto T^{-\nu }$ and since the current density scale
as $J\propto kT/\xi $, the crossover to nonlinear behavior sets in at $%
J_{nl}\propto T^{1+\nu }$ which depends strongly on the yet unknown
critical exponent $\nu $. Also, the linear resistivity $R_{L} $ is finite at
any nonzero temperature but thermally activated, $R_{L}\propto \exp
(-E_{b}/kT)$. Thus the relaxation time $\tau \propto 1/R_{L}$ diverges
exponentially for decreasing temperatures. We can then consider the behavior
of the dimensionless ratio $E/JR_{L}$ which must satisfy the scaling form 
\cite{hyman}

\begin{equation}
E/JR_{L}=g(J/T^{1+\nu })  \label{scalt0}
\end{equation}
if the assumption of a zero-temperature transition is correct. In Fig. 4(b)
we show the scaling plot according to Eq. (\ref{scalt0}) for the
lowest temperatures and current densities which verifies the
scaling assumption and provides an estimate of $\nu =2.1(2)$ and an energy
barrier $E_{b}=1.2$. This value of $\nu$ is consistent with the estimate 
$\sim 1.9$  based on the previous finite-size scaling of defect energy in 
the ground state \cite{bgsg}. 
Similar analysis at different dilution $x=0.45$ gives $\nu =2.2(2)$ 
and $E_{b}=0.9$ and at different frustration $f=1/4$ gives $\nu
=2.3(2)$ and $E_{b}=1.45$. We note that our  estimate of $\nu \sim 2$ 
is roughly the same as the value obtained for the gauge-glass model of 
strongly disordered two-dimensional superconductors \cite{hyman,eg98}
which may suggest a common universality class. However, it should be noted
that, for $f=1/2$, the system has a global reflection symmetry 
($\theta_i \rightarrow - \theta_i$) in addition to the  rotational
symmetry \cite{villain} and one would expect, similarly to the
XY (chiral) spin glass \cite{kawamura} which shares the same feature, 
two different divergent 
correlation lengths $\xi_s$ and $\xi_c$ with  corresponding  distinct exponents $\nu_s$
and $\nu_c$. 
In fact, for the chiral glass model a different universality class with $\nu_s \sim 1$
has been found from a current-voltage scaling analysis \cite{eg98}. 
On the other hand, an analytic study \cite{hilhorst} of the XY spin glass 
for a particular  distribution of disorder find a common exponent. Our estimates
suggest that this could also be the case for the present percolative type of
disorder or else the exponents are too close to be resolved within the
accuracy of our estimate. An apparent common universality class of vortex glass models
with clearly distinct symmetries has also been found in three dimensions \cite{by}.
In addition, close to the percolation threshold $x_{p}$,
the above scaling analysis  based on a single diverging length scale is not valid,
as  one must also take into account the percolation correlation length $\xi _{p}$ 
and the fractal nature of the system at smaller length scales \cite{martinoli,affolter}.

In summary, we have studied the interplay of phase coherence and
vortex-glass state in two-dimensional diluted Josephson-junction arrays with
average rational values of frustration. For $f=1/2$, we found  evidence of a phase
coherence threshold value $x_{s}$ much below the geometric percolation
threshold $x_{p}$. This is in contrast with the conclusions of 
Gupta and Teitel \cite{gt99} for a Josephson-junction array with positional
disorder where no phase coherence is expected at finite temperatures even for small
disorder at length scales much larger than a disorder dependent length. 
Further work is required to verify whether the present study 
only reflects the finite-length scale of the system sizes used in the calculation
or is a consequence of different type of disorder. In addition, since $f=1/2$ has 
a particular reflection symmetry, which is preserved in presence of random dilution, 
the behavior for other values of $f$ could be qualitatively different.  
On the other hand, experiments \cite{harris} are often done on systems sizes comparable to 
our largest system size and thus the current-voltage scaling behavior discussed here 
should be observable. In the range  $x_{s}<x<x_{p}$ the array behaves as a
zero-temperature vortex glass with activated nonzero linear resistance at
finite temperatures and critical currents much less sensitive to variations
in $f$ than in the phase coherent region. Our results suggest that the phase
coherence threshold can be identified experimentally as the change in the
transport properties from the weak to the  strong disorder regime. However, 
the numerical estimates of critical quantities from the current-voltage scaling  analysis 
should be regarded as rough magnitudes which can be measured experimentally and used to 
verify the prediction of a change in behavior of the transport 
properties in the different phases. Equilibrium simulations are required to confirm the 
observed critical behavior and obtain more accurate estimates of the critical exponents. 

\medskip We have benefited from many discussions with M. Benakli, S.R.
Shenoy and J. Affolter. The work of E.G. was supported by FAPESP and of 
D.D. by CONICET and Fundaci\'{o}n Antorchas (Proy.
A-13532/1-96). We also acknowledge the ICTP (Trieste) where this work
was started.

\newpage

\begin{figure}[btp]
\centering\epsfig{file=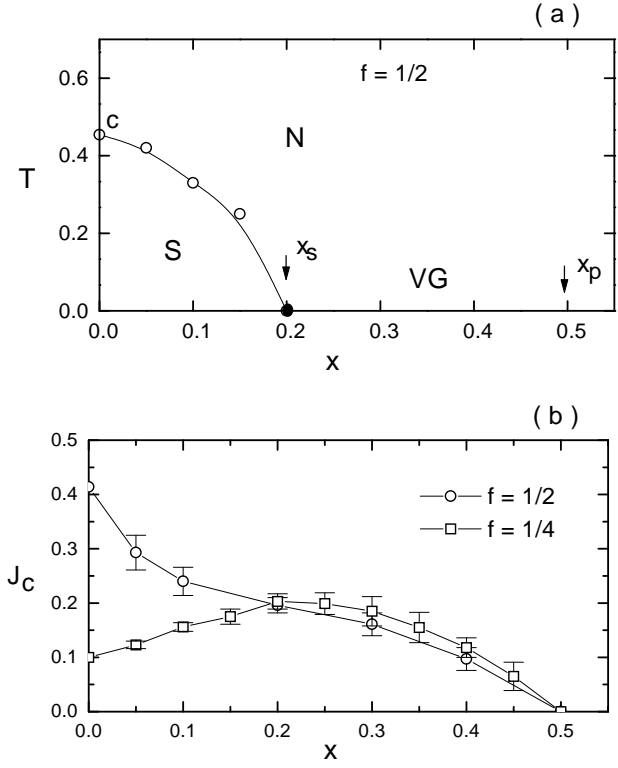,bbllx=1cm,bblly=0.5cm,bburx=20cm,
 bbury=26cm,width=8.5cm}
 
\caption{(a) Phase diagram of a diluted Josephson-junction array as a
function of temperature $T$ and concentration $x$ of diluted junctions, for
an average rational frustration $f=1/2$. The superconducting phase is
denoted by S, the normal phase by N and the short-range vortex glass state
by VG. The geometrical percolation threshold is indicated by $x_p$ and the
phase-coherence threshold by $x_s$. (b) Critical current densities $J_c$ as
a function of dilution $x$ for different values of frustration $f$. Critical
temperatures (open circles) in (a) were obtained from current-voltage
scaling analysis and the phase coherence threshold $x_s$ (filled circle) was
inferred from the change in the behavior of $J_c(f)$ in (b). }
\label{fig1}
\end{figure}

\begin{figure}[btp]
\centering\epsfig{file=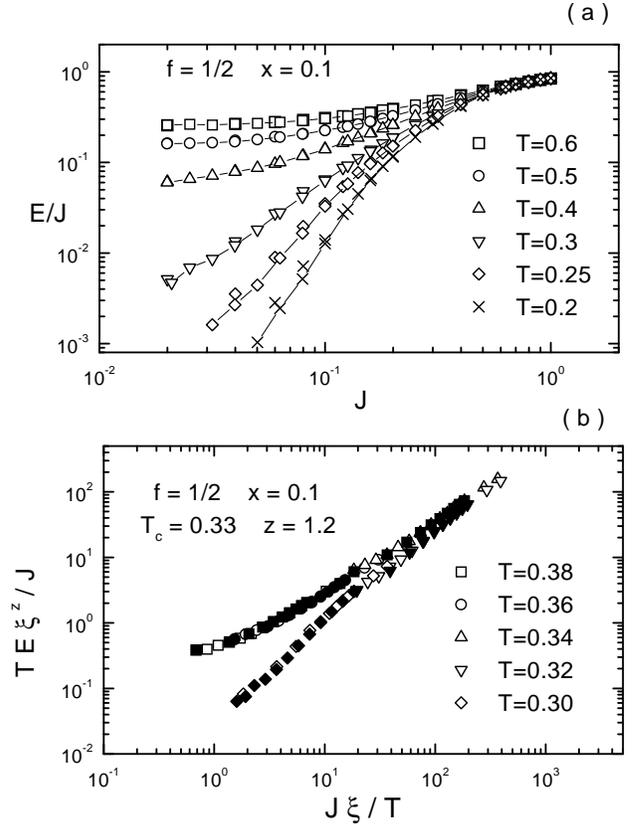,bbllx=1cm,bblly=0.5cm,bburx=20cm,
 bbury=28cm,width=8.5cm}

\caption{ (a) Nonlinear resistivity $E/J$ as a function of temperature for a
dilution $x=0.1$ below the phase coherence threshold $x_s$ and system sizes
$L=64$ and $L=128$ (symbols connected by lines). 
(b)Scaling plot of the data (not indicated in (a)) 
for the smallest range near $T_c$ and smallest current densities. 
Open symbols correspond to $L=64$ and filled ones to $L=128$. }
\label{fig2}
\end{figure}

\begin{figure}[btp]
\centering\epsfig{file=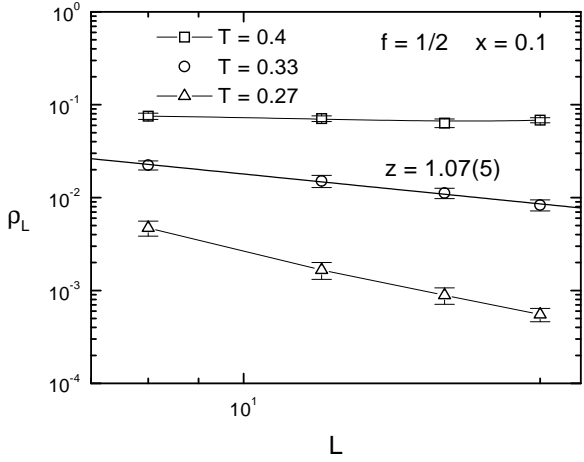,bbllx=1cm,bblly=3cm,bburx=20cm,
 bbury=28cm,width=8.5cm}

\caption{ Linear resistance as a  function of system size $L$ for different
temperatures at $x=0.1$. A power-law fit at $T_c=0.33$ gives an estimate
of the dynamical exponent $z$.}
\label{fig3}
\end{figure}

\begin{figure}[btp]
\centering\epsfig{file=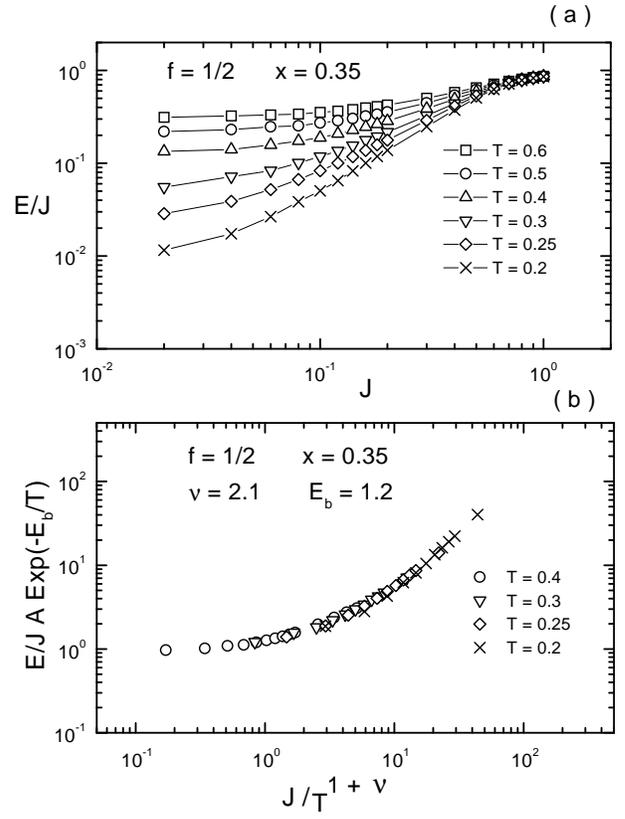,bbllx=1cm,bblly=0.5cm,bburx=20cm,
 bbury=26cm,width=8.5cm}

\caption{(a) Nonlinear resistivity $E/J$ as a function of temperature for a
dilution $x=0.35$ above the phase coherence threshold $x_s$. (b) Scaling
plot of the data in (a) for the lowest temperatures and current densities
according to a $T=0$ transition. }
\label{fig4}
\end{figure}

\end{document}